\documentclass[12pt]{iopart}
\usepackage{verbatim}
\usepackage{bm}
\usepackage{iopams}  
\usepackage{bm}
\usepackage{mathrsfs}
\usepackage{enumerate}
\usepackage{verbatim}
\usepackage{xcolor}
\usepackage{stmaryrd}
\usepackage{url}
\usepackage{graphicx}	
\usepackage{subfigure}
\usepackage[perpage]{footmisc}

\newcommand{\eqref}[1]{(\ref{#1})}

\newcommand{\erfc}{\mathrm{erfc}}

\newcommand{\Id}{\mathbf{I}}

\newcommand{\R}{\mathbf{R}}

\newcommand{\rot}{\mathsf{SO}(3)}

\newcommand{\FN}{F_N}
\newcommand{\FNA}{\ave{\FN}}
\newcommand{\free}{\mathscr{F}}

\newcommand{\pot}{\mathscr{U}}

\newcommand{\body}{\mathcal{B}}

\newcommand{\expo}[1]{\mathrm{e}^{#1}}
\newcommand{\dd}{\mathrm{d}}
\newcommand{\q}{q}

\newcommand{\qq}{\bm{q}}

\newcommand{\x}{\bm{x}}
\newcommand{\av}{\bm{a}}
\newcommand{\dens}{\rho}
\newcommand{\dist}{\varrho}

\newcommand{\ave}[1]{\left\langle{#1}\right\rangle}
\newcommand{\ceil}[1]{\left\lceil{#1}\right\rceil}
\newcommand{\floor}[1]{\left\lfloor{#1}\right\rfloor}

\newcommand{\conf}{\Omega}
\newcommand{\meas}{|\conf|}
\newcommand{\Conf}{\mathcal{Q}}
\newcommand{\denso}{\rho_0}

\newcommand{\Fij}{\Phi_{ij}}

\newcommand{\dq}{\dd\q_1\dots \dd\q_N}

\newcommand{\ori}{\omega}

\newcommand{\evo}{v_\mathrm{e}}
\newcommand{\Evo}{V_\mathrm{e}}
\newcommand{\ZN}{Z_N}
\newcommand{\GN}{G_N}
\newcommand{\GNA}{\ave{G_N}}
\newcommand{\fA}{\ave{f}}

\newcommand{\GGN}{\mathcal{G}_N}
\newcommand{\FFN}{\mathcal{F}_N}
\newcommand{\GGnu}{\mathcal{G}^{(1)}_\nu}
\newcommand{\Tnu}{\mathcal{T}_\nu}
\newcommand{\Fnu}{\mathcal{F}_\nu}
\newcommand{\graph}{G}
\newcommand{\tree}{T}
\newcommand{\forest}{F}

\newcommand{\freeO}{\free_\mathrm{O}}

\begin{document}

\title{Onsager's missing steps retraced}
\author{Peter Palffy-Muhoray$^1$, Epifanio G Virga$^2$\footnote{\emph{On leave} from Department of Mathematics, University of Pavia, via Ferrata 5, 27100 Pavia, Italy. E-mail: \texttt{eg.virga@unipv.it}} and Xiaoyu Zheng$^3$} 
\address{$^1$ Liquid Crystal Institute, Kent State University, OH, USA}
\ead{mpalffy@kent.edu}
\address{$^2$ Mathematical Institute, University of Oxford, UK}
\ead{virga@maths.ox.ac.uk}	
\address{$^3$ Department of Mathematical Sciences, Kent State University, OH, USA}
\ead{zheng@math.kent.edu}

\vspace{10pt}
\begin{indented}
\item[]\today
\end{indented}

\begin{abstract}
Onsager's paper on phase transition and phase coexistence in anisotropic colloidal systems is a landmark in the theory of lyotropic liquid crystals. However, an uncompromising scrutiny of Onsager's original derivation reveals that it would be rigorously valid only for ludicrous values of the system's number density (of the order of the reciprocal of the number of particles).  Based on Penrose's tree identity and an appropriate variant of the mean-field approach for purely repulsive, hard-core interactions, our theory shows that Onsager's theory is indeed valid for a reasonable range of densities.
\end{abstract}

\vspace{2pc}

\submitto{\JPCM}
%\maketitle
% 
% For two-column output uncomment the next line and choose [10pt] rather than %[12pt] in the \documentclass declaration

\section{Introduction}\label{sec:intro}
The most successful theory of the formation of a nematic liquid crystal phase
due to steric interactions is Onsager's celebrated work on the effects of
shape on the interaction of colloidal particles \cite{onsager:effects}. The model
successfully describes not only the nematic-isotropic
transition \cite{samborski:isotropic}, but also the phase separation into an isotropic and
nematic phase \cite{lekkerkerker:understanding}.

The motivation of this paper is the desire to understand the success of Onsager's classical theory for liquid crystals based on purely repulsive, hard-core interactions. This theory, appropriately credited with being the first incarnation of modern density functional theory, was highly influenced by Mayer's theory for real gases and by the cluster expansion formalism \cite{mayer:statistical}. However, when this putative parental connection is   scrutinized, as done in some detail in Sect.~\ref{sec:Onsager_steps}, one readily realizes that the system's number densities that would rigorously grant compatibility for the two theories are ludicrously small, of the order of $1/N$, where $N$ is the total number of particles. This at least is the disheartening conclusion one arrives at by following closely and uncompromisingly the steps of Onsager's original derivation.

Then, where does the success of Onsager's theory come from? Here we attempt to answer this question by retracing the steps that were missed in Onsager's derivation. Our answer will not be trivial, but we hope that it will be reassuring: Onsager's theory is well rooted, its domain of validity is limited, but meaningful.

The major mathematical tool of our inquiry will be Penrose's tree identity \cite{penrose:convergence}, combined with a mean-field approach to purely steric interactions within the canonical ensemble. In this perspective, Onsager's theory, though unchanged in content, will emerge in a different light. This should not be a surprise: of Onsager we could say what Brydges \cite{brydges:mayer} said of Mayer: ``He was a pioneer and his derivations twist and turn''.  

This is the structure of the paper. In Sect.~\ref{sec:Onsager_steps}, we reproduce and criticize Onsager's original derivation.  In Sect.~\ref{sec:free_energy}, we set the scene where our developments are placed by introducing the Helmholtz free energy in a canonical ensemble of  \emph{complex} particles with multiple degrees of freedom. Sect.~\ref{sec:Penrose_identity} is devoted to Penrose's tree identity and its relevance to our estimate of the free energy. In Sect.~\ref{sec:mean_field}, we combine Penrose's identity with a variant of the mean-field approximation that we find suitable for repulsive, hard-core interactions, which are at the heart of our theory. We shall also arrive there at an explicit and reasonable bound on the number density that marks the range of validity of Onsager's theory. In Sect.~\ref{sec:Onsager_approximation}, we connect our approach with the density functional theory for liquid crystals, of which Onsager's model is perhaps the most rightful ancestor. Finally, in Sect.~\ref{sec:conclusions}, we collect the conclusions of our paper and comment on the possible avenues that it opens for the study of dense nematic liquid crystals, well above the range of validity of Onsager's theory.

The paper is closed by an appendix, which is rather technical in nature: it presents the major mathematical tool employed in this paper, namely, the graph-combinatoric, asymptotic  estimates of Britikov~\cite{britikov:asymtotic} that afford corroborating the validity of Onsager's conclusion, if not his derivation.

\section{Onsager's missing steps}\label{sec:Onsager_steps}
In this section, we condense our critique of Onsager's seminal paper \cite{onsager:effects}. We adopt Onsager's own point of view, even  his language and notation whenever possible. [We refer to his equation (xx) in \cite{onsager:effects} as (Oxx)].

Onsager's model expresses the irreducible Mayer cluster integrals \cite{mayer:statistical}
$\beta_{1}$ and $\beta_{2}$ in terms of particle shape parameters, and
estimates the partition function $B(T)$ using these. The free energy is then readily obtained from the partition function;
the equilibrium state of the system is that which minimizes the free energy.

Onsager's model has been widely used, with volume fractions up to and in
excess of $0.70$ (see, for example, \cite{van_roij:transverse}). Onsager himself expressed the hope that the
results would describe ``fairly well'' isotropic solutions with volume fractions up to
$0.25$. We shall thus focus on
the range of densities for which his free energy expression can be
expected to hold.

Onsager's derivation proceeds as follows.
The number of states (measured in units of generalized volume) are given by
(O1)
\begin{equation}
B(T)=\frac{1}{N!}\int \expo{-\frac{u}{kT}}\dd\tau,\label{E1}
\end{equation}
where $u$ stands for the total interaction potential and $\dd\tau$ is the element of
volume in configuration space.
The resulting Helmholtz free energy is given by
\begin{equation}
F=-kT\ln B.\label{E1.5}
\end{equation}
For repulsive forces, the hard-particle pair-interaction energy is (O17)
\begin{equation}
w=\sum_{i<j}w_{ij},
\end{equation}
where
\begin{equation}
w_{ij}=w_{2}(q_{i},q_{j}),
\end{equation}
and for simplicity, we ignore orientation and regard the $q_{i}$s' as only
coordinates of position.\ The Mayer function is defined as in (O18) by
\begin{equation}
\Phi_{ij}=\expo{-\frac{w_{ij}}{kT}}-1.
\end{equation}
We note here that $\Phi_{ij}\leqq0$.

Expanding the integrand of (\ref{E1})
gives (O19),
\begin{equation}
\expo{-\frac{w}{kT}}=1+\sum_{i<j}\Phi_{ij}+\sum_{i<j,i^{\prime}<j^{\prime}}%
\Phi_{ij}\Phi_{i^{\prime}j^{\prime}}+\cdots.
\end{equation}
Substitution into the integrand of (\ref{E1}) and rearranging terms produces
\begin{eqnarray}
I&=\int \expo{-\frac{u}{kT}}\dd\tau\nonumber\\
&=\int\Big(1+\sum_{i<j}\Phi_{ij}+\sum_{i<j<k}\Phi
_{ij}\Phi_{jk}\Phi_{ki}+\cdots\Big)\dd\tau_{1}\dd\tau_{2}\cdots \dd\tau_{N},
\end{eqnarray}
where we have kept, in accordance with \cite{onsager:effects}, the lowest order  terms
(according to the number of variables $\tau_{i}$ which appear in the
integrand). We note that, by its very definition, $I\geqq0$. Integration gives%
\begin{eqnarray}
I=V^{N}&+V^{N-2}\sum_{i<j}\int\int\Phi_{ij}\dd\tau_{i}\dd\tau_{j}\nonumber\\
&+V^{N-3}
\sum_{i<j<k}\int\int\int\Phi_{ij}\Phi_{jk}\Phi_{ki}\dd\tau_{i}\dd\tau_{j}\dd\tau
_{k}+\cdots.
\end{eqnarray}
Now, omitting the higher order terms (not shown) and noting that $w_{ij}$ only
depends on the difference of particle coordinates, since there are
$N(N-1)/2\simeq N^{2}/2$ pairs and $N(N-1)(N-2)/3!\simeq N^{3}/6$ triplets, we
have,
\begin{eqnarray}
%\fl
I=V^{N}\Bigg(1 
&
+\frac{N^{2}}{2V^{2}}\int\int\Phi_{12}\dd\tau_{1}\dd\tau_{2}
\nonumber\\
&
+\frac{N^{3}
}{6V^{3}}\int\int\int\Phi_{12}\Phi_{23}\Phi_{31}\dd\tau_{1}\dd\tau_{2}\dd\tau
_{3}\Bigg).\label{E2}%
\end{eqnarray}
Using the cluster integral notation (O20), in (\ref{E2}) we can set
\begin{equation}
\beta_{1}=\frac{1}{V}\int\int\Phi_{12}\dd\tau_{1}\dd\tau_{2},
\end{equation}
with the units of volume, and%
\begin{equation}
\beta_{2}=\frac{1}{2V}\int\int\int\Phi_{12}\Phi_{23}\Phi_{31}\dd\tau_{1}%
\dd\tau_{2}\dd\tau_{3},
\end{equation}
with the units of volume squared.\ Since $\Phi_{ij}\leqq0$, it follows that
$\beta_{1}\leqq0$ and $\beta_{2}\leqq0$.
In terms of these, we have
\begin{equation}
I=V^{N}\left(1+\frac{N^{2}}{2V}\beta_{1}+\frac{N^{3}}{3V^{2}}\beta_{2}\right).\label{3}%
\end{equation}

We now consider $\beta_{1}$ and $\beta_{2}$ in more detail.
Since $w_{1j}$ depends only on the difference of coordinates, we can integrate
over $\tau_{1}$, and obtain%
\begin{equation}
\beta_{1}=\int\Phi_{12}d\tau_{12},
\end{equation}
where $\tau_{12}$ indicates the difference of the coordinates of particles $1$
and $2$.\ It is clear that $\beta_{1}$ is just the excluded volume of
particles $1$ and $2$; we therefore write
\begin{equation}\label{eq:E14}
\beta_{1}=-av_{0},
\end{equation}
where $v_{0}$ is the volume of one particle, and note that, for spherical
particles in $3D$, $a=8$.
Similarly, for $\beta_{2}$, we can integrate out $\tau_{1}$ and obtain%
\begin{equation}
\beta_{2}=\frac{1}{2}\int\int\Phi_{12}\Phi_{23}\Phi_{31}\dd\tau_{12}\dd\tau_{13},
\end{equation}
which we write as
\begin{equation}\label{eq:E16}
\beta_{2}=-bv_{0}^{2}.
\end{equation}
For spherical particles $b=15$.
We then write equation (\ref{3}) in terms of $a$ and $b$, and obtain
\begin{equation}
I=V^{N}\left(1-\frac{N^{2}}{2V}av_{0}-\frac{N^{3}}{3V^{2}}bv_{0}^{2}\right).\label{3.5}%\
\end{equation}
Defining the volume fraction $\phi$ as
\begin{equation}
\phi=\frac{Nv_{0}}{V},
\end{equation}
we rewrite \eqref{3.5} as
\begin{equation}
I=V^{N}\left(1-\frac{1}{2}aN\phi-\frac{1}{3}bN\phi^{2}\right),\label{3.6}%
\end{equation}
where $I$ must be non-negative. Assuming that $\phi$ is small, we neglect the
last term, and note that non-negativity requires that
\begin{equation}
\phi<\frac{2}{aN}.\label{E4}%
\end{equation}
(If the last term is not neglected, the constraint on $\phi$ is even stronger.)

This is our key criticism. Since the number of states must be non-negative,
Onsager's approach requires that the volume fraction occupied by particles be
of the order of $1/N$.
Given that even modest systems have $N\approx10^{20}$ particles, clearly the
region of validity of the model is so small as to be of little practical
value. We shall return to equation (\ref{3.6}) and (\ref{E4}) subsequently.

Ignoring the constraint on the volume fraction, substitution of 
(\ref{3.5}) into (\ref{E1}) and (\ref{E1.5}), gives
\begin{equation}
\ln B=N+N\ln\left(\frac{V}{N}\right)+\ln\left(1-\frac{1}{2}aN\phi-\frac{1}{3}bN\phi
^{2}\right),\label{5}%
\end{equation}
where also Stirling's approximation has been used.
Assuming that $\frac{1}{2}aN\phi+\frac{1}{3}bN\phi^{2}\ll1$ and expanding the
logarithm in Taylor's series, we obtain
\begin{equation}
\ln B=N\left[1+\ln\left(\frac{V}{N}\right)+\frac{1}{2}\beta_{1}\frac{N}{V}+\frac{1}{3}\beta_{2}\left(\frac{N}{V}\right)^{2}\right],\label{7}
\end{equation}
after having substituted for $\phi$ in \eqref{5} and made use of both \eqref{eq:E14} and \eqref{eq:E16}. Equation \eqref{7} is precisely (O21).

It seems therefore that, although not at all obvious from (\ref{7}) and
not pointed out in his paper, Onsager's expression for the free energy is
valid only in the regime where the volume fraction of particles is no more
than of order $1/N$, as prescribed by the \emph{hidden}, but stringent requirement of positivity for $I$ at the chosen level of approximation. If this region of validity is observed, the theory is
essentially useless for any practical system.

One might be temped to argue that the positivity of $I$ in (\ref{3.6})
may be maintained by the terms which have been neglected; indeed, this would
have to be the case. However, these terms cannot then be neglected as they
have been in (\ref{5}) and (\ref{7}).

Alternately, one might be tempted to argue, as in \cite{frenkel:onsager}, that
the theory is valid in the limit as the aspect ratio of particles goes to
infinity, in which case terms of order greater than $2$ may indeed be
neglected. However, even in that case, the constraint in
(\ref{E4}), which is on density rather than shape, must still hold.

Historically, Onsager's starting point \eqref{7} has naively been regarded as a truncation of the density expansion  (13.43) in Mayer and Mayer's book \cite{mayer:statistical}. That too, however, had not been fully justified before Onsager's times.\footnote{The proof given in \cite{mayer:statistical} is based on the estimate of the logarithm of a sum with a large number of terms by the logarithm of the maximum term. The Mayers duly warn the reader that this is acceptable only  when all terms in the sum are positive, which unfortunately is not the case here. They refer to the work of Born and Fuchs \cite{born-fuchs:statistical} for a rigorous proof of the very same statement, which does not call on that restrictive (and unverified) assumption. However, a critical reading of Born and Fuchs' paper reveals a number of technical assumptions that must be verified for the result to be proven rigorously: one is the hypothesis about the existence of a single maximum point in their application of the method of steepest descent to estimate the partition function; another assumption is stated in their inequality (3.12), which involves a whole range of functions defined as sums of series (which one fails to see how can be justified in practical terms). The first rigorous proof of the convergence of the Mayer expansion in the canonical ensemble came only a few years ago \cite{pulvirenti:cluster}.} 

Onsager's beautiful and seminal theory of the isotropic-nematic transition is
in impressive agreement with experiment and simulation; it must therefore be
either correct for realistic densities, or at least nearly correct. However,
justification for the theory in this regime is absent in \cite{onsager:effects}. We hope to provide one here, independent of either Mayers' expansion and its convergence.

\section{Helmholtz free energy}\label{sec:free_energy}
We consider a classical system of $N$ particles, for which $\x_i$, $i=1,\dots,N$, denote the positions in space of the individual centres of mass and $\omega_i\in\conf$ represents the $i$th  particle's generalized \emph{orientation}. In general, we think of $\conf$ as an $m$-dimensional space, on which the action of the rotation group $\rot$ in ordinary three-dimensional space is specified by a group $\mathsf{G}$ of mappings with $\mathsf{G}\ni g_\R:\conf\to\conf$ corresponding to the rotation $\R\in\rot$, and enjoying the property $g_{\R_1\R_2}=g_{\R_1}g_{\R_2}$, for all rotations $\R_1$ and $\R_2$. We shall denote by $\dd\x$ the volume measure in three-dimensional space, and by $\dd\omega$ a $\mathsf{G}$-invariant measure over $\conf$, for which $|\conf|$ is the total measure. 

$\conf$ is the \emph{orientation space}. For rigid particles, which represent the most common form of colloids, we can identify $\conf$ with $\rot$; the appropriate measure on $\conf$ is then the Haar measure on $\rot$, when this latter is viewed as a topological group, and if we identify $\omega$ with the triple $(\vartheta,\varphi,\psi)$ of Euler's angles then $\dd\omega$ is most conveniently expressed as 
\begin{equation}\label{eq:rotation_measure}
\dd\omega=\sin\vartheta\dd\vartheta\dd\varphi\dd\psi
\end{equation} 
and $|\conf|=8\pi^2$.

In the following, whenever needed, we shall denote by $\q_i=(\x_i,\omega_i)$ the \emph{configuration} of the $i$th particle and by $\qq=(\q_1,\dots,\q_N)$ is the \emph{collective} configuration of all particles. For a given rotation $\R\in\rot$, we shall denote $\R\qq:=(\R\q_1,\dots,\R\q_N)$, where $\R\q_i:=(\R\x_1,g_\R(\omega_i))$. Likewise, for any translation $\av$, we shall write $\av\qq:=(\av\q_1,\dots,\av\q_N)$, where $\av\q_i:=(\x_i+\av,\omega_i)$. To shorten our formul\ae, we shall also write $\dd\q_i=\dd\x_i\dd\omega_i$ and, occasionally, $\dd\qq=\dq$. We shall further assume that all particles are confined within a region $\body$  of volume $V$ in three-dimensional space, so that $\Conf:=\body\times\conf$ will be their \emph{configuration space}. Similar general treatments of systems of \emph{complex} particles can also be found in \cite{poghosyan:abstract} and \cite{jansen:multispecies}.\footnote{For liquid crystal molecular theories, this formalism was also adopted in \cite{gartland:minimum} and Chap.~1 of \cite{sonnet:dissipative}.}

In a given collective configuration $\qq\in\Conf^N$, particles interact through a pair-wise potential $U=U(\q_i,\q_j)$, which has the following properties:\footnote{As we learn from standard textbooks, such as \cite{ruelle:statistical} and \cite{gallavotti:statistical}, these  in general  guarantee a good statistical mechanical behaviour of the system. We therefore assume them, though they would not be strictly needed for the applications to hard-core colloidal particles that we envision here.}
\begin{enumerate}
\item 
$U$ is \emph{symmetric}: $U(\q_i,\q_j)=U(\q_j,\q_i)\quad\forall\ \q_i\neq\q_j$;
\item  $U$ is \emph{invariant} under \emph{translations}: $U(\av\q_i,\av\q_j)=U(\q_i,\q_j)\quad\forall\ \av,\ \q_i\neq\q_j$;
\item $U$ is \emph{invariant} under \emph{rotations}: $U(\R\q_i,\R\q_j)=U(\q_i,\q_j)\quad\forall\ \R\in\rot,\ \q_i\neq\q_j$;
\item $U$ is \emph{stable}, i.e., there is a constant $b\geqq0$ such that 
\begin{equation}\label{eq:total_potential}
\pot(\qq):=\sum_{1\leqq i<j\leqq N}U(\q_i\,\q_j)\geqq-Nb.
\end{equation}
\end{enumerate}

Hard-core potentials between extended, \emph{deformable} particles are encompassed in this class: for them, $U(\q_i,\q_j)=+\infty$ whenever the pairs $(\x_i,\omega_i)$ and $(\x_j,\omega_j)$ are such that particle $i$ and particle $j$ would partially overlap in space.  We are especially interested in \emph{steric} interactions, for which $U(\q_i,\q_j)$ is either $+\infty$ or $0$, depending on whether particles in configurations $\q_i$ and $\q_j$ \emph{do} overlap or \emph{not}, respectively. Clearly, for these interactions, $b=0$ in \eqref{eq:total_potential}.

The purely configurational Helmholtz free energy is defined  as
\begin{equation}\label{eq:free_energy_N_definition}
\FN=-kT\ln\ZN,
\end{equation}
where $k$ is Boltzmann's constant, $T$ is the absolute temperature, and $\ZN$ is the \emph{canonical} partition function,
\begin{equation}\label{eq:canonical_partition_N_function}
\ZN:=\frac{1}{N!}\GN.
\end{equation}
Here
\begin{equation}\label{eq:number_of_states}
\GN:=\int_{\Conf^N}\expo{-\frac{1}{kT}\pot(\qq)}\dd\qq
\end{equation}
and $\pot$ is the total potential defined in \eqref{eq:total_potential}. In \eqref{eq:canonical_partition_N_function}, the factor $1/N!$ accounts for the  indistinguishability of particles. As is customary, kinetic terms have been omitted in \eqref{eq:number_of_states}, under the assumption that they contribute quadratically to the Hamiltonian and so can be integrated exactly and then disposed of as coefficients of the integral of states in   \eqref{eq:number_of_states}; these coefficients depend only on the temperature and shift the free energy in \eqref{eq:free_energy_N_definition} by an inessential additive constant. The case of steric interactions between \emph{rigid particles} falls in the category for which this assumption is valid.

Following Mayer~\cite{mayer:statistical_I}, we introduce the definition 
\begin{equation}\label{eq:Mayer_function}
\Fij:=\Phi(\q_i,\q_j):=\expo{-\frac{1}{kT}U(\q_i,\q_j)}-1
\end{equation}
and rewrite \eqref{eq:number_of_states} in the following equivalent form
\begin{equation}\label{eq:number_of_states_Mayer}
\GN=\int_{\Conf^N}\prod_{1\leqq i<j\leqq N}(1+\Fij)\dq.
\end{equation}
The Mayer function $\Fij$ is particularly meaningful in the case of steric interactions, since then its values are either $0$, when particles $i$ and $j$ do not overlap, or $-1$, when they do; for these interactions, $\Fij$ is the \emph{anti-characteristic} function of the overlapping set (in configuration space) for two particles. Clearly, if any two rigid particles overlap, it follows from \eqref{eq:number_of_states_Mayer} that that configuration does not contribute to the integral of states $\GN$. The introduction of Mayer's functions is an ingenious device that allows the extension of the integral in \eqref{eq:number_of_states_Mayer} to the whole available space without overcounting the number of admissible states.

As already done by Mayer and his collaborators in  a well-known series of papers ~\cite{mayer:statistical_I, mayer:statistical_II,mayer:statistical_III,mayer:statistical_IV,mayer:statistical_V,mayer:statistical_VI}, following an earlier suggestion of Ursell~\cite{ursell:evaluation}, (see also the textbook \cite{mayer:statistical}, for a systematic exposition of this method\footnote{According to \cite{stell:cluster}, Yvon~\cite{yvon:theorie} was another precursor of Mayer in writing cluster expansions. He is also credited with the stipulation that cluster integrals in the more general case of a non-uniform system should be identical in their topological form with the cluster expansions valid in the uniform case. This  conclusion was implicit in the iterative procedure on which his method is based. For this extension of the cluster  expansion formalism, which will also be recalled later in Sect.~\ref{sec:Onsager_approximation} of this paper, see  \cite{morita:new,stillinger:equilibrium,de_dominicis:variational}, whose general ideas can be retraced in Chapter 5 of the book \cite{massignon:mecanique}. A later extension, but along different lines, was proposed in \cite{rowlinson:virial} (see also \cite{masters:virial} for a broad review). All these formulations are in the grand-canonical ensemble, whereas here we are only interested in the canonical ensemble counterpart, which is far less common.}) we give \eqref{eq:number_of_states_Mayer} yet another equivalent form, phrased in the language of graphs:
\begin{equation}\label{eq:number_of_states_graphs}
\GN=\int_{\Conf^N}\sum_{\graph\in\GGN}\prod_{(i,j)\in\graph}\Fij\dq,
\end{equation}
where $\GGN$ is the collection of all (undirected) graphs on $N$ labelled vertices and the pair $(i,j)$ denotes the edge connecting vertices $i$ and $j$ in a graph $\graph$. Here vertices represent particles and edges their interactions. Again, in the paradigmatic case of steric interactions, an edge links any two overlapping particles, whose shared space is inaccessible in reality and must be subtracted from the total count of configurational states. In this view, the Mayer function $\Fij$ becomes an \emph{edge function} defined on all possible graphs of $\GGN$.

It should be noticed that $\GGN$ also contains the \emph{empty graph}, which has no edges, but only vertices: when in \eqref{eq:number_of_states_graphs} $\graph$ is the empty graph, the product over the non-existing edges is conventionally set equal to $1$ to warrant accord between \eqref{eq:number_of_states_Mayer} and \eqref{eq:number_of_states_graphs}.

We shall see in the following section how an identity valid for general edge functions reduces \eqref{eq:number_of_states_graphs} to a form that readily suggests an approximate evaluation of $\GN$, appropriate if only graphs with a limited connectivity could be selected from $\GGN$.

\section{Penrose's tree identity}\label{sec:Penrose_identity}
A general, quite unexpected identity proved by Penrose~\cite{penrose:convergence} can be used to write \eqref{eq:number_of_states_graphs} in an equivalent and rather suggestive form. In recent years, Penrose's identity has been variously re-interpreted and extended \cite{sokal:bounds,procacci:convergence}. Here, for completeness, we shall present its original formulation, after recalling some elementary graph terminology.

A graph $\graph$ is said to be \emph{connected} if for any two vertices in $\graph$ there is \emph{at least} one path of edges linking them. The collection of all connected graphs on $\nu$ vertices will be denoted by $\GGnu$. A graph $\tree\in\GGnu$ is called a \emph{tree} if for any two vertices in $\tree$ there is \emph{precisely} one path of edges linking them. The collection of all trees on $\nu$ vertices will be denoted by $\Tnu$. A \emph{forest} is a disjoint union of trees. The collection of all forests on $\nu$ vertices will be denotes by $\Fnu$.

Penrose's identity is based on a \emph{reduction} algorithm, often also called a \emph{partition scheme}, which associates a tree $\tree\in\Tnu$ with any connected graph $\graph\in\GGnu$ and, when reversed, starting from any tree $\tree\in\Tnu$, it identifies the \emph{maximal} graph $\tree^\ast\in\GGnu$ that reduces to $\tree$ under the direct partition scheme. 

\begin{figure}
	\centering
    \subfigure[]{\includegraphics[width=0.32\linewidth]{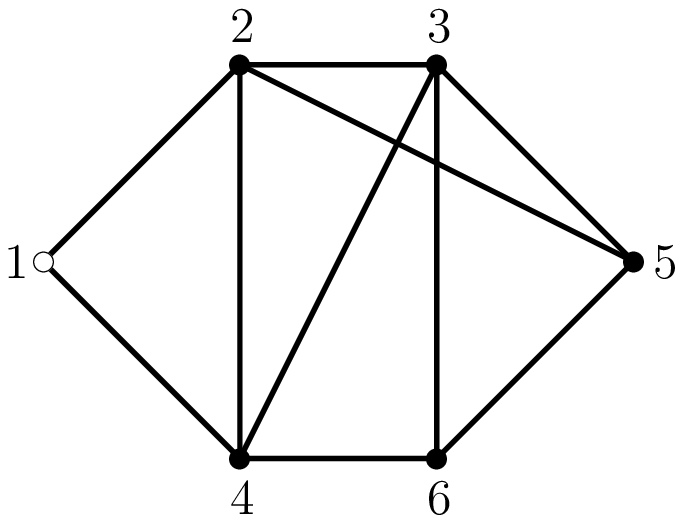}}
	%\hspace{.01\linewidth}
	\subfigure[]{\includegraphics[width=0.32\linewidth]{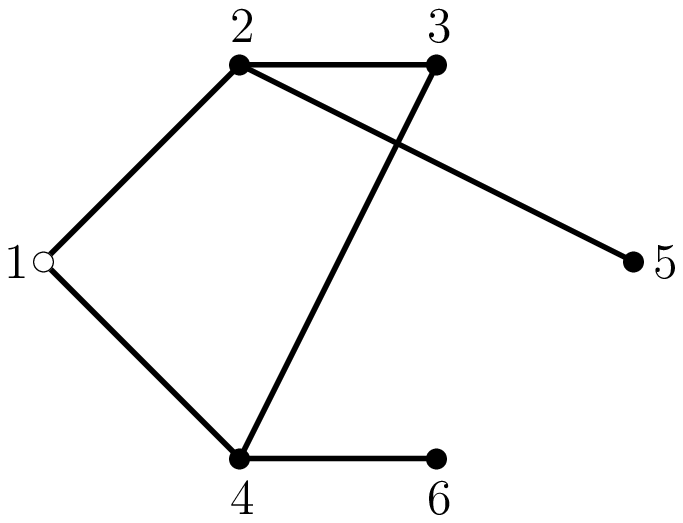}}
	%\hspace{.01\linewidth}
	\subfigure[]{\includegraphics[width=0.32\linewidth]{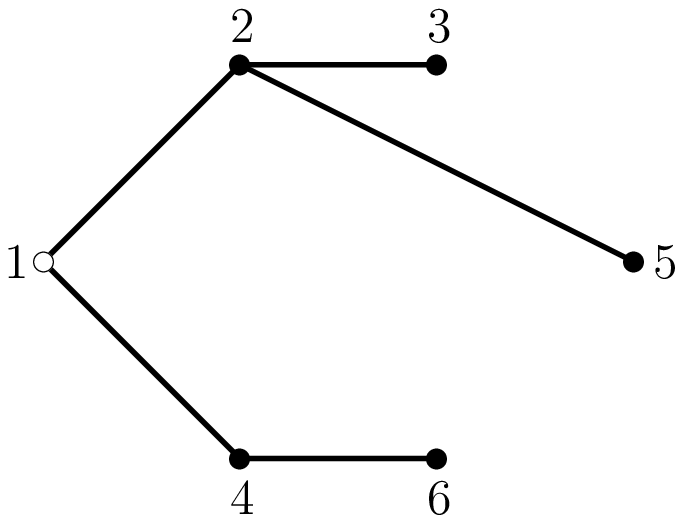}}
	\caption{Penrose's partition scheme applied to a graph $\graph\in\GGnu$. %\\
	(a) A connected graph with $6$ vertices; their weights are $w_1=0$, $w_2=w_4=1$, $w_3 =w_5 =w_6 =2$. %\\
    (b) The graph obtained after one iteration of the partition scheme. %\\
    (c) The tree $\tree$ obtained after a second iteration of the partition scheme.}
	\label{fig:Penrose_descending}%
\end{figure}

We shall not reproduce Penrose's proof here (see \cite{virga:course}, for more pedagogical details), but for completeness we shall illustrate Penrose's partition scheme in a simple case. This scheme consists in the following steps:
\begin{enumerate}[(a)]
	\item Choose one vertex in a \emph{connected} graph $\graph$, say vertex $1$;
	\item Assign a \emph{weight}  to every vertex $i\neq1$, defined as the number $w_i$ of edges in the shortest path linking $i$ to $1$ in $\graph$ (by definition, $w_1=0$);
	\item Delete all edges between vertices of equal weight;
	\item For every vertex $i\neq1$, delete all edges joining vertex $i$ to a vertex with weight $w_i-1$, apart from the one that among these has the least index. 
\end{enumerate}
Each of these steps leaves the weight $w_i$ unchanged. The result of the construction is independent of the choice of vertex $1$. By repeatedly applying this scheme, any graph $\graph\in\GGnu$ is reduced to a tree $T\in\Tnu$. Figure~\ref{fig:Penrose_descending} shows a simple example of this reduction.

\begin{figure}
	\centering
	\subfigure[]{\includegraphics[width=0.32\linewidth]{Penrose_T.eps}}
	%\hspace{.01\linewidth}
	\subfigure[]{\includegraphics[width=0.32\linewidth]{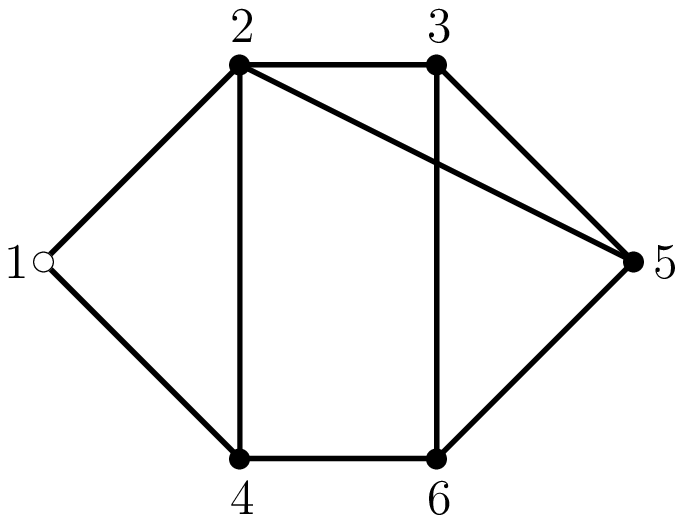}}
	%\hspace{.01\linewidth}
	\subfigure[]{\includegraphics[width=0.32\linewidth]{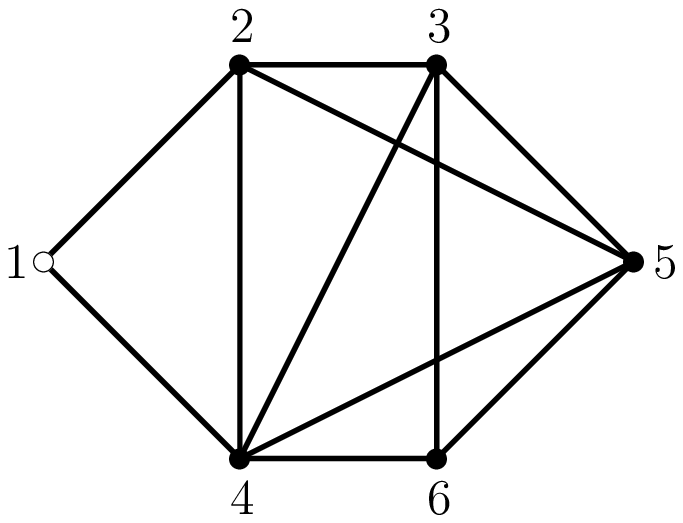}}
	\caption{Construction of the maximal graph $\tree^\ast$ from a given  tree $\tree$ through Penrose's reverse partition scheme. %\\
		(a) The tree $\tree$ obtained in Fig.~\ref{fig:Penrose_descending}(c) from the graph $\graph$ in Fig.~\ref{fig:Penrose_descending}(a) through Penrose's partition scheme. The weights of the vertices are $w_1=0$, $w_2=w_4=1$, $w_3 =w_5 =w_6 =2$. %\\
		(b) The graph obtained after one iteration of the reverse partition scheme. %\\
		(c) The maximal graph $\tree^\ast$ obtained after a second iteration of the reverse partition scheme.}
	\label{fig:Penrose_ascending}%
\end{figure}

To construct the maximal graph $\tree^\ast$ originated by a tree $\tree$, Penrose's \emph{reverse} scheme prescribes the following steps: 
\begin{enumerate}[(a$^\prime$)]
	\item Choose one vertex in a \emph{tree} $\tree$, say vertex $1$;
	\item Assign a weight $w_i$ to every vertex $i$, as in the direct scheme above;
	\item Join all pairs of vertices with the same weight;
	\item Join every vertex $i\neq1$ of weight $w_i$ to all vertices of weight $w_i-1$ with index greater than the largest index of the vertices of weight $w_i-1$ to which is already joined in $\tree$. 
\end{enumerate}
These steps, precisely as the direct steps (a)--(d) illustrated above, leave the weights $w_i$ unchanged and produce a graph independent of the choice of vertex $1$. By repeated application of these steps to a tree $\tree$, we obtain the maximal graph $\tree^\ast$ that Penrose's direct partition scheme reduces to $\tree$. Figure~\ref{fig:Penrose_ascending} shows the application of the reverse construction to the tree $\tree$ in Fig.~\ref{fig:Penrose_descending}(c); as expected in general, the maximal graph $\tree^\ast$ in Fig.~\ref{fig:Penrose_ascending}(c) is different from the parent graph of $\tree$ in Fig.~\ref{fig:Penrose_descending}(a).

\begin{figure}
	\centering
	\includegraphics[width=0.2\linewidth]{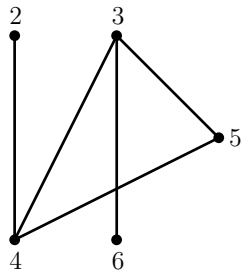}
	\caption{The graph difference $\tree^\ast\setminus\tree$ for the simple example in Fig.~\ref{fig:Penrose_ascending}.}
	\label{fig:graph_difference}%
\end{figure}

We are now in a position to state Penrose's tree identity. Given an edge function $\Phi$ defined on all connected graphs $\graph$ on $\nu$ vertices, 
\begin{equation}\label{eq:Penrose_tree_identity}
	\sum_{\graph\in\mathcal{G}^{(1)}_{\nu}}\prod_{(i,j)\in\graph}\Phi_{ij}= \sum_{T\in\mathcal{T}_{\nu}}\prod_{(i,j)\in T}\Phi_{ij}\prod_{(h,k)\in T^\ast\setminus T}(1+\Phi_{hk}).
\end{equation}
In  a less formal, but perhaps more effective language,  identity \eqref{eq:Penrose_tree_identity} says that ``the sum over all connected graphs on $\nu$ vertices is equal to the sum over just tree graphs on $\nu$ vertices with more complicated weights'' \cite{brydges:mayer}. 
In \eqref{eq:Penrose_tree_identity}, such correcting weights are expressed as products over the edges of the graph difference $\tree^\ast\setminus\tree$, which for the simple examples in Fig.~\ref{fig:Penrose_ascending} is shown in Fig.~\ref{fig:graph_difference};
simpler expressions can be given by use of a clever hard-sphere construction in two higher spatial dimensions, as shown in \cite{brydges:branched}. It is worth remarking that for purely steric interactions 
\begin{equation}\label{eq:correcting_weights_bounds}
0\leqq1+\Phi_{hk}\leqq1\quad\forall\ h\neq k,
\end{equation}
and so all correcting weights in \eqref{eq:Penrose_tree_identity} are bounded in the same interval.

Equation \eqref{eq:Penrose_tree_identity} is valid for connected graphs, but it is easily extended to all graphs on $N$ vertices in $\GGN$, which are of interest here. Let $\FFN$ denote the collection of all forests on $N$ vertices. For any forest $\forest\in\FFN$, we construct the maximal graph $\forest^\ast\in\GGN$ reducible to $\forest$ by applying Penrose's reverse scheme to all connected components of $\forest$, that is, to all trees of $\forest$, and taking the union of the maximal graphs $\tree_l^\ast$ obtained from each individual tree $\tree_l\subset\forest$. Thus the integral of states in \eqref{eq:number_of_states_graphs} can be rewritten as
\begin{equation}\label{eq:number_of_states_forests}
	\GN=\int_{\Conf^N}\sum_{\forest\in\FFN}\prod_{(i,j)\in\forest}\Phi_{ij} \prod_{(h,k)\in\forest^\ast\setminus\forest}(1+\Phi_{hk})\dd\q_1\dots \dd\q_N,
\end{equation}
where the edge function $\Phi$ is now meant to be the Mayer function  \eqref{eq:Mayer_function}. 

In the following section, making use of \eqref{eq:number_of_states_forests}, we shall introduce a mean-field approximation to $\GN$, which will be proved to be equivalent to Onsager's approximation to the free energy of colloidal orientable particles, within a range of validity which can be quantitatively estimated in a general fashion.

\section{Mean-field approximation}\label{sec:mean_field}
Though exact, formula \eqref{eq:number_of_states_forests} still offers formidable difficulties to the explicit computation of the total number of states $\GN$, even in the simplest case, that of a gas of hard spheres. For this reason, we resort to a \emph{mean-field} approximation, which at this level of detail is embodied by replacing $\Phi$ in \eqref{eq:number_of_states_forests} with its average $\ave{\Phi}$ over $\Conf^2$. The reader should heed that with this approximation we lose all information concerning particle orientations and that the resulting free energy will thus be that of an isotropic phase. To recover this lost information, in Sect.~\ref{sec:Onsager_approximation}, we shall revisit Onsager's original ``mixture argument'',  which we still find compelling. 

We find it more convenient to introduce the function $f:=-\Phi$, which for steric interactions represents the \emph{characteristic function}  of the overlapping set in $\Conf$ of two particles. Thus, setting $\fA=-\ave{\Phi}$, we replace $\GN$ with its mean-field approximation $\GNA$, for which all integrals in \eqref{eq:number_of_states_forests} are immediately evaluated:
\begin{equation}\label{eq:number_of_states_non_trees}
\GNA:=V^N\meas^N\sum_{\forest\in\FN}(-1)^{e(\forest)}\fA^{e(\forest)}(1-\fA)^{e(\forest^*\setminus\forest)},
\end{equation}
where $e(\graph)$ denotes the number of edges in a graph $\graph$. Though immensely simpler than \eqref{eq:number_of_states_forests}, even \eqref{eq:number_of_states_non_trees} is not explicitly computable in terms of $\fA$ and $N$. Before introducing a further and possibly drastic approximation in the mean-field formula for $\GNA$, which will be shown to be equivalent to Onsager's unjustified approximation, we comment briefly on the physical meaning of $\fA$.

For a pair of particles at positions $\x$ and $\x'$ with  configurations $\ori$ and $\ori'$ interacting through a potential $U(\x,\ori;\x',\ori')$ enjoying the properties described in Sect.~\ref{sec:free_energy}, $\fA$ is formally defined as
\begin{equation}\label{eq:average_f_definition}
\fA:=\frac{1}{V^2\meas^2}\int_{\Conf^2}f(\x,\ori;\x',\ori')\dd\x\dd\ori\dd\x'\dd\ori',
\end{equation}
where 
\begin{equation}\label{eq:f_definition}
f(\x,\ori;\x',\ori'):=1-\expo{-\frac{1}{kT}U(\x,\ori;\x',\ori')}.
\end{equation}
Since
\begin{equation}
f(\x+\av,\ori;\x'+\av,\ori')=f(\x,\ori;\x',\ori')
\end{equation}	
for all translations $\av$, assuming that $\body$ is a macroscopic domain of volume $V$ and that the range of $f$ (where it does not vanish) is much smaller than the linear size of $\body$, as is certainly the case for purely steric interactions, we may write $\fA$ in \eqref{eq:f_definition} as
\begin{equation}\label{eq:f_Mayer_form}
\fA=\frac{1}{V\meas^2}\int_{\Delta\body\times\conf^2}M(\bm{r};\ori,\ori')\dd\bm{r}\dd\ori\dd\ori',
\end{equation}	
where $\Delta\body:=\body-\body$ is the Minkowski difference\footnote{The Minkowski difference $\mathcal{A}-\mathcal{B}$ between sets of points $\mathcal{A}$ and $\mathcal{B}$ in a Euclidean space represented as  vectors issued from a given origin is the set of all possible differences $\bm{a}-\bm{b}$ of vectors $\bm{a}\in\mathcal{A}$ and $\bm{b}\in\mathcal{B}$. Thus, for example, for $\body$ a ball of radius $R$, $\body-\body$ is a ball of radius $2R$.} between $\body$ and itself, and the function $M$ is defined as
\begin{equation}\label{eq:M_definition}
M(\bm{r};\ori,\ori'):=f(\bm{0},\ori;\bm{r},\ori'),
\end{equation}
having assumed that the origin $\bm{0}$ belongs to $\body$. Said differently, \eqref{eq:f_Mayer_form} follows from assuming that the fluid is \emph{uniform}, so that any particle can be taken as the origin of coordinates.

For purely steric interactions $M$ takes only values $1$ or $0$; we say that it is the characteristic function of the \emph{Mayer region} in $\Delta\body\times\conf^2$. \Eref{eq:f_Mayer_form} clearly reveals $\fA$ as the excluded volume \emph{fraction}. For \emph{rigid particles}, $\conf$ can be identified with the rotation group $\rot$ and the invariance under rotations of $U$ entails that 
\begin{equation}
\fA=\frac{1}{8\pi^2 V}\int_{\Delta\body\times\rot}M_\mathrm{R}(\bm{r},\R)\dd\bm{r}\dd\R,
\end{equation}
where the \emph{reduced} Mayer function $M_\mathrm{R}$ depends only on the relative displacement $\bm{r}$ and the relative rotation $\R$ of two particles. Formally,
\begin{equation}
M_\mathrm{R}:=M(\bm{r};\Id,\R),
\end{equation}
where $\Id$ is the identity in $\rot$.

Our further approximation, admittedly severe, but meaningful---as we shall see, neglects the excluded volume fraction $\fA$ in the non-tree component of  \eqref{eq:number_of_states_non_trees}, where we simply replace $1-\fA$ with $1$:
\begin{equation}\label{eq:mean_field_sum_number_of_states_forests}
\GNA\approx V^N\meas^N\sum_{\forest\in\FN}(-1)^{e(\forest)}\fA^{e(\forest)}.
\end{equation}
It is clear what we renounce with this approximation for $\GNA$ and where this can go badly wrong: \eqref{eq:number_of_states_non_trees} shows that as $\fA$ grows the admissible states become scant, but this is not captured by \eqref{eq:mean_field_sum_number_of_states_forests}. Qualitatively, this latter should become inadequate as the number density
\begin{equation}\label{eq:number_density_definition}
\denso:=\frac{N}{V}
\end{equation}
increases; our approximation should possess the character  of a mean-field for \emph{dilute} systems. We shall see below that our method is capable of determining quantitatively the bound for $\denso$ within which the approximation in \eqref{eq:mean_field_sum_number_of_states_forests} is expected to be valid. One may also wonder whether \eqref{eq:mean_field_sum_number_of_states_forests} can be interpreted in terms of neglected correlations. Properly speaking, no bound is imposed on the order of correlations being considered: it is more the topology of correlations than their order to be limited. Said differently, the permitted level of correlation among interacting particles does not allow for closed loops in their connectivity network, but it is otherwise unlimited. 

Properly speaking, $\GNA$ in \eqref{eq:mean_field_sum_number_of_states_forests} should \emph{not} be called an \emph{approximation} for the real number of states $\GN$ because we can offer no estimate of the error $|\GN-\GNA|$; rather, at best $\GNA$ \emph{mimics} $\GN$ instead of approximating it.

No matter how simple \eqref{eq:mean_field_sum_number_of_states_forests} may look, computing explicitly the sum over all forests $\forest$ on $N$ vertices is not a trivial matter. We shall accomplish this task by rearranging the terms in \eqref{eq:mean_field_sum_number_of_states_forests} and estimating this sum in the limit of large $N$. Such an estimate will be performed with a view on obtaining the mean-field free energy per particle, which according to \eqref{eq:free_energy_N_definition} and \eqref{eq:canonical_partition_N_function} is defined as
\begin{equation}\label{eq:mean_field_free_energy_per_particle}
\frac{1}{N}\FNA:=-\frac{kT}{N}\ln\frac{1}{N!}\GNA.
\end{equation} 
This will be especially instrumental to the purpose of computing the thermodynamic limit, where both $N$ and $V$ diverge, while keeping $\denso$ fixed.

We first rewrite \eqref{eq:mean_field_sum_number_of_states_forests} by grouping all forests $\forest$ on $N$ vertices by the number $n$ of edges they feature (alternatively, they could be grouped by the number $K=N-n$ of trees they possess):
\begin{equation}\label{eq:mean_field_sum_number_of_states}
\GNA\approx V^N\meas^N\sum_{n=0}^{N-1}(-1)^n C(N,n)\fA^n,
\end{equation}
where $C(N,n)$ is the cardinality of the collection $\FFN^{(n)}\subset\FFN$ of all forests with $n$ edges on $N$ vertices. In the limit of large $N$, $C(N,n)$ was variously estimated by Britikov~\cite{britikov:asymtotic}.\footnote{Britikov's estimates play in our theory the same role that  combinatoric estimates, proved in a series of papers \cite{riddell:theory,ford:combinatorial_I,ford:combinatorial_II,ford:combinatorial_III,ford:combinatorial_IV,uhlenbeck:theory} all co-authored by Uhlenbeck, played in Mayer's theory.} Full details on Britikov's estimates are given in \ref{sec:Britikov}. It is proved there that
\begin{equation}\label{eq:mean_field_number_of_states}
\GNA\approx V^N\meas^N\expo{-Nx}\quad\mathrm{for}\quad N\to\infty,
\end{equation}
where $x$, which is defined as
\begin{equation}\label{eq:x_definition}
x:=\case12N\fA,
\end{equation}
is a quantity of order $\Or(1)$ in the thermodynamic limit. However, equation \eqref{eq:mean_field_number_of_states}, which will soon lend itself to the retracing of Onsager's missing steps, is valid only for $0<x<x_0$, where  (see \ref{sec:Britikov})
\begin{equation}\label{eq:x_0}
x_0:=\case12W\left(\case1\rme\right)\doteq0.139,
\end{equation}
and $W$ is the Lambert function (\cite{NIST:DLMF}, \S\,4.13.10).

For rigid particles with \emph{average} pair-excluded volume $\evo$, 
\begin{equation}\label{eq:v_e_introduction}
x=\case12\denso\evo,
\end{equation}
and the requirement for the validity of \eqref{eq:mean_field_number_of_states} reads as
\begin{equation}\label{eq:density_upper_bound}
\denso \evo<W\left(\case{1}{\rme}\right)\doteq0.278.
\end{equation}
This upper bound for the density is to be compared with the least upper bound found by Lebowitz and Penrose~\cite{lebowitz:convergence} to ensure convergence of the virial series (for the pressure in terms of density) within Mayer's theory in the grand-canonical ensemble. In the case of steric interactions, Lebowitz and Penrose's sufficient condition for the convergence of the virial series reads as\footnote{Lebowitz and Penrose's condition has recently been extended by Procacci and Yuhjtman~\cite{procacci:convergence}, but for steric interactions their extended condition reproduces that of Lebowitz and Penrose. The latter was also shown to be sufficient for the convergence of the virial series in the canonical ensemble by Pulvirenti and Tsagkarogiannis~\cite{pulvirenti:cluster} and for the analyticity of the Helmholtz free energy as a function of density in the canonical ensemble by Morais and Procacci~\cite{morais:continuous}. Expressing the upper bound in \eqref{eq:Lebowitz_Penrose_condition} in terms of the Lambert function $W$ was suggested in \cite{tate:virial}.}
\begin{equation}\label{eq:Lebowitz_Penrose_condition}
\denso \evo<\max_{0<w<1}w\left(2\expo{-w}-1\right)=\frac{\left[W\left(\case{\rme}{2}\right)-1\right]^2}{W\left(\case{\rme}{2}\right)}\doteq0.145.
\end{equation}
The bound in \eqref{eq:density_upper_bound} was derived to ensure validity of our approximate evaluation of $\GNA$ in the canonical ensemble, and so it has no connection with the bound in \eqref{eq:Lebowitz_Penrose_condition} which was derived to ensure the analyticity of the equation of state in the grand-canonical ensemble. Though dissimilar in  origin, \eqref{eq:density_upper_bound} and \eqref{eq:Lebowitz_Penrose_condition} are both similar requirements of diluteness. 

Combining \eqref{eq:mean_field_number_of_states} and \eqref{eq:mean_field_free_energy_per_particle}, by use of \eqref{eq:x_definition}
and the classical Stirling's approximation, we arrive at the following formula for the free energy per particle in the mean-field approximation,
\begin{equation}\label{eq:free_energy_per_particle}
\frac{1}{NkT}\FNA=\ln\denso-\ln\meas-1+\case12N\fA,
\end{equation}
which readily reduces to
\begin{equation}\label{eq:free_energy_per_volume}
\frac{1}{\meas VkT}\FNA=\frac{\denso}{\meas}\ln\frac{\denso}{\meas}-\frac{\denso}{\meas}+\frac12\left(\frac{\denso}{\meas}\right)^2\fA\meas V,
\end{equation}
to express the free energy per unit volume for a uniform system. Computing the Legendre transform of $\FNA/V$ as a function of $\denso$ delivers the pressure $P$ as
\begin{equation}\label{eq:pressure}
\frac{P}{kT\denso}=1+\case12\denso\fA V.
\end{equation}
For rigid particles such as hard spheres \eqref{eq:pressure} takes the more familiar form
\begin{equation}\label{eq:pressure_familiar}
\frac{P}{kT\denso}=1+\case12\denso\evo,
\end{equation}
which is the van der Waals equation for a purely repulsive interaction potential. In our theory, all these formul\ae\ are subject to the same bound on density \eqref{eq:density_upper_bound} that makes \eqref{eq:mean_field_number_of_states} valid.

Though \eqref{eq:pressure} and \eqref{eq:pressure_familiar} have the appearance of Mayer's virial series truncated at the first correction to the ideal gas pressure, here they have a completely different justification: they are originated from neglecting all non-tree graphs in \eqref{eq:number_of_states_non_trees} and summing over all possible forests of trees in \eqref{eq:mean_field_sum_number_of_states_forests}. No hypothesis on the existence of a virial series was ever made, and so no truncation has occurred in \eqref{eq:pressure} or \eqref{eq:pressure_familiar}. Instead, the term linear in $\denso$ featuring there results from a laborious re-summation on all trees in the average total number of states. Had we been able to include graphs containing cycles in \eqref{eq:number_of_states_non_trees} (and to count properly them; combinatorics is  the main obstruction here),  we would have further corrected the ideal gas  equation of state by adding a function of $\denso$, which need \emph{not} be a power of the density.

There is one special instance where   \eqref{eq:pressure} and \eqref{eq:pressure_familiar} are exact: as shown in \cite{frisch:classical,frisch:errata} for a fluid of hard-spheres, this is the limiting case of infinite space dimensionality (see also \cite{luban:comment} and \cite{frisch:reply} for the debate that this hypothetical limit has sparked). The proof of \eqref{eq:pressure_familiar} in \cite{frisch:classical} is phrased in the traditional framework of Mayer's cluster expansion: it was shown that all higher order correction terms in the virial expansions vanish in the high dimensionality limit. In the language of our paper, this result is not surprising, since in infinite spatial dimensions all graphs entering  \eqref{eq:number_of_states_non_trees} are bound to be forests: loosely stated, in infinite dimensions tentative cycles \emph{miss}  closure most of the times.

In our approach, firmly rooted in three-dimensional space, \eqref{eq:mean_field_number_of_states} fails to hold for $x>x_0$ because Britikov's estimates show that the average total number of states $\GNA$, as given by \eqref{eq:mean_field_sum_number_of_states}, then alternates in sign as $N$ increases, in contrast with its very meaning. This is a failure that we ascribe to the \emph{minimal connectivity} hypothesis, as we call the approximation that led us to  \eqref{eq:mean_field_sum_number_of_states_forests}, since it neglects  all the cycles present in the maximal graphs $\forest^*$ generated from forests by Penrose's reverse algorithm. Such a minimal connectivity approximation is bound to break down as the density increases; the merit of Britikov's estimates is to have made explicit the upper bound $x_0$ of $x$ above which this approximation cannot be trusted. We shall indicate in the following section how this approximation is related to Onsager's  free-energy functional for liquid crystals.

\section{Onsager's approximation in density-functional form }\label{sec:Onsager_approximation}
Here, building upon \eqref{eq:free_energy_per_particle}, which Onsager~\cite{onsager:effects} had taken with little, or no comment from Mayer and Mayer's book \cite{mayer:statistical} (see Sect.~\ref{sec:Onsager_steps} for a discussion of such \emph{missing steps} as they emerge from a critical reading of Onsager's paper),\footnote{Mayer and Mayer's book \cite{mayer:statistical} had been published  just a couple of years before Onsager's first announcement of his theory  \cite{onsager:anisotropic}.} we shall arrive at Onsager's celebrated functional for systems of rigid, anisomeric particles \emph{non-uniformly} distributed in orientation space $\conf$. This is perhaps the first manifestation of the general approach that is now known as the \emph{density functional} theory.

Onsager's original derivation is ingenious; rigid  particles are regarded to be distributed in their orientation as if they were members of different coexisting species. This is the essence of the famous \emph{multi-species} argument that can be retraced in textbooks (see also \cite{virga:course} for an exposition that parallels closely Onsager's original paper). Rather than reproducing here the classical derivation, we shall present a synthetic, formal argument, which could be extended to more general settings.  

We start from assuming that particles are distributed in configuration space $\Conf$ with a density $\dens$, which obeys the normalization constraint
\begin{equation}\label{eq:N_normalization}
\int_\Conf\dens(\q)\dd\q=N.
\end{equation} 
With this in mind, we regard \eqref{eq:free_energy_per_volume} as providing the \emph{local} free energy per unit volume in $\Conf$. Thus the \emph{total} free energy functional $\free[\dens]$ associated with the distribution density $\dens$ will follow by integrating the local free-energy density over $\Conf$. While the integration of the first two terms in the right side of \eqref{eq:free_energy_per_volume} is rather unambiguous, the integration of the last term requires discernment. This is a term quadratic in the density where the extra volume $\meas V$ entails a further integration. Combining these formal prescriptions with \eqref{eq:average_f_definition}, we justify the following \emph{definition} for $\free$,
\begin{equation}\label{eq:free_energy_functional_definition}
\frac{1}{kT}\free[\dens]:=\int_\Conf\dens(\q)\left(\ln\dens(\q)-1\right)\dd\q+\frac12\int_{\Conf^2}\dens(\q)\dens(\q')f(\q,\q')\dd\q\dd\q',
\end{equation}
which is the standard form of density functional theory \cite{wu:density}. An alternative and possibly more convincing derivation is given in \cite{palffy:effects}, where \eqref{eq:free_energy_functional_definition} is obtained through a partition of $\Conf$ in interacting cells individually described by \eqref{eq:free_energy_per_volume}. In both ways, we replace the \emph{unweighted} average of (minus) the Mayer function in equation \eqref{eq:free_energy_per_volume} with its $\rho(q)$-weighted average, thus reintroducing the information on particle orientations that had been lost in the mean-field approximation.  

The functional in \eqref{eq:free_energy_functional_definition} is more general than that employed  by Onsager, which was intended for \emph{homogeneous} systems of rigid particles  interacting only via hard-core repulsion. We imagine  such particles to be distributed in orientation space $\conf$, which here we identify with $\rot$, through a probability density $\dist$, which is normalized to unity,
\begin{equation}\label{eq:1_normalization}
\int_\conf\dist(\ori)\dd\ori=1.
\end{equation}
Therefore we represent the density $\dens$ as
\begin{equation}\label{eq:density_representation}
\dens(\q)=\denso\dist(\ori),
\end{equation}
where $\denso$ is the number density in \eqref{eq:number_density_definition}. By letting $\dens$ in \eqref{eq:free_energy_functional_definition}  be expressed as in \eqref{eq:density_representation} and making use of \eqref{eq:N_normalization} and  \eqref{eq:1_normalization}, we readily arrive at a functional of $\dist$ expressing the dimensionless free energy per particle as
\begin{eqnarray}
\fl
\frac{1}{NkT}\free[\dist]=\ln\denso-1&+\int_\conf\dist(\ori)\ln\dist(\ori)\dd\ori\nonumber\\
&+\frac12\denso\int_{\conf^2}\dist(\ori)\dist(\ori')\Evo(\ori,\ori')\dd\ori\dd\ori'=:\freeO[\denso;\dist]\label{eq:Onsager_free_energy_functional},
\end{eqnarray}
where $\Evo(\ori,\ori')$ is the pair-excluded volume defined as
\begin{equation}\label{eq:pair_excluded_volume_definition}
\Evo(\ori,\ori'):=\int_{\Delta\body}M(\bm{r};\ori,\ori')\dd\bm{r}
\end{equation}
and $M$ is the characteristic function of the Mayer region defined by \eqref{eq:M_definition}. The average pair-excluded volume $\evo$ introduced in \eqref{eq:v_e_introduction} is thus related to $\Evo$ through 
\begin{equation}\label{eq:v_e_formula}
\evo=\frac{1}{\meas^2}\int_{\conf^2}\Evo(\ori,\ori')\dd\ori\dd\ori'.
\end{equation} 
$\freeO$ is Onsager's free-energy functional for colloidal systems: it is a functional of the distribution density $\dist$ whose minimizers, representing equilibria, depend on the number density $\denso$, which features as a bifurcation  parameter.

The more general functional $\free$ in \eqref{eq:free_energy_functional_definition} has so far lived, as it were, two independent lives. One life springs from the extension of Mayer's cluster expansions to non-homogeneous systems first obtained by Stillinger and Buff~\cite{stillinger:equilibrium}, building on earlier work of Yvon~\cite{yvon:theorie} and Morita and Hiroike~\cite{morita:new}. In this approach, \eqref{eq:free_energy_functional_definition} is simply viewed as a truncated functional in the same spirit as \eqref{eq:pressure} is seen (but not by us) as a truncated virial expansion.

The other life springs from the density functional theory originally put forward by Hohenberg and Kohn~\cite{hohenberg:inhomogeneous} to describe the ground state of an electron gas and further extended to encompass atomic fluids by Evans~\cite{evans:nature}, the fluid-solid transition and interfaces by Ramakrishnan and Yussouff~\cite{ramakrishnan:first} followed by Haymet and Oxtoby~\cite{haymet:molecular}, and liquid crystals by Stecki and Kloczkowski~\cite{stecki:stability} followed by Sluckin and Shukla~\cite{sluckin:molecular}.\footnote{All these contributions to the density functional theory were formulated in the grand-canonical ensemble. An interesting formulation of the theory in the canonical ensemble has recently been proposed in \cite{dwandaru:variational}, building on a two-step minimization method of Levy~\cite{levy:universal}.} As lucidly phrased in \cite{sluckin:molecular}, ``[w]e may regard the theory expounded in this paper as a \emph{mesoscopic} theory of the isotropic-nematic transition. In spirit it lies between a fully microscopic theory---in which the liquid partition function is evaluated, and a \emph{macroscopic} theory---which depends totally on a  number of phenomenological parameters.'' A central role is played in this theory by a functional of the form \eqref{eq:free_energy_functional_definition}, where $f(\x,\ori;\x',\ori')$ is replaced by the direct correlation function  (first introduced by Orstein and Zernike~\cite{ornstein:accidental}), which in Onsager's case simply reduces to the pair-excluded volume $\Evo(\ori,\ori')$. In this approach, the Mayer function is actually the lowest-order approximation to the direct correlation function in a density-functional expansion of the free energy for a non-uniform fluid (as shown, for example in Sects.~3.8 and 3.9 of \cite{harsen:theory}).

In our approach, we aspire to give \eqref{eq:free_energy_functional_definition} a third life: we see it as mean-field mimic of the free energy generated by a minimal connectivity microscopic network.\footnote{Perhaps not surprisingly, it was proved that \eqref{eq:free_energy_functional_definition} is exact in the limit of infinite space dimensions not only for hard-spheres \cite{frisch:nonuniform}, but also for liquid crystals \cite{carmesin:liquid}.} It is still a mesoscopic theory, but with a clear microscopic basis and a well-defined range of validity.
	
%\section{Onsager's density functional}\label{sec:Onsager_functional}
\section{Conclusion}\label{sec:conclusions}
Although we were sceptical of the rigour  of Onsager's original derivation of his celebrated model for lyotropic liquid crystals, we were confident in its  validity, which had been confirmed (at least qualitatively) by both experiments and simulations. Yet the fundamental questions remained as to why and in what range Onsager's model is valid. This paper offers an answer to both issues.

The validity of Onsager's model does \emph{not} rest on the neglect of all but the first irreducible cluster integral of Mayer's theory. Rather, it rests on the (admittedly complicated) resummation of all possible forests of tree graphs that contribute to the total number of states. Surprisingly, the net result is  the same, but the theoretical significance is quite different. It is still an approximation, but its range of validity, given explicitly by the upper bound in \eqref{eq:x_0}, is small but reasonable. In this approximation, correlations are not quantitatively limited, only their topology is.

Onsager's bifurcation analysis of the minimizers of the free-energy functional $\freeO$ in  \eqref{eq:Onsager_free_energy_functional} for rigid rods was performed within a one-parameter family of distribution functions $\dist$. In his analysis, he introduced a dimensionless concentration $c$, which turns out to be precisely our $x$ in \eqref{eq:v_e_introduction}. The critical value $x_\mathrm{c}$ for the first-order isotropic-nematic transition is $x_\mathrm{c}\doteq3.681$ and the isotropic phase encounters a transcritical bifurcation at $x=x_\mathrm{t}\doteq4$ (see also \cite{virga:course}). These values were confirmed by the combined analytic-numerical strategy adopted in \cite{kayser:bifurcation} and by a deeper bifurcation analysis recently performed in \cite{vollmer:critical}. Both $x_\mathrm{c}$ and $x_\mathrm{t}$ fall considerably above the upper bound $x_0\doteq0.139$ in \eqref{eq:x_0}. So a leap of faith is still required to trust quantitatively the results of Onsager's analysis; we have shown that it is at least qualitatively justifiable.

In graph theoretic language, cycles are to be added in the minimal connectivity strategy, which here only includes trees, to improve the approximate evaluation of  the total number of states. They might be classified, for example, as are Husimi trees \cite{husimi:note}, of which the (ordinary) trees considered in this paper are special cases. Once higher order Husimi trees are considered, the functional $\freeO$ is expected to change, and not necessarily by supplementing the quadratic form in $\dist$ now in \eqref{eq:Onsager_free_energy_functional} with a cubic one, but we have no clue yet as to what form this functional may have.

Pursuing our interest in the behaviour of dense nematic liquid crystals, we have recently taken a completely different avenue \cite{nascimento:density}: instead of building the average total number of states $\GNA$ through higher levels of connectivity, we have estimated it by recursion on the number of particles $N$. That approach and the one presented in this paper are completely different, we hope they will be shown to be complementary.   

\ack
We are grateful to P. Hohenberg and R. Petschek for inspiring discussions on the subject of this paper. Moreover, E.G.V. would like to express his gratitude to D. Tsagkarogiannis for having introduced him to the charming intricacies of cluster expansions, to D. Brydges for his learned and suggestive  encouragements to pursue the theoretical approach presented here, to A. Procacci for explaining the subtleties of Penrose's tree identity, and to O. Penrose himself for his patience and thoughtfulness in listening to a preliminary presentation of the contents of this paper while he was visiting Oxford.
Finally, E.G.V. acknowledges the kind hospitality of the Oxford Centre
for Nonlinear PDE, where this work was completed while
he was visiting the Mathematical Institute at the University of
Oxford. This manuscript has benefited from the comments and suggestions of two Reviewers who have contributed (especially one of them) to improve considerably the presentation of our work. We are grateful to both.

\appendix

\section{Britikov's estimates}\label{sec:Britikov}
Our estimate for $\GNA$ in Sect.~\ref{sec:mean_field} relies heavily on a paper by Britikov~\cite{britikov:asymtotic} that proves various asymptotic estimates for the number $C(N,n)$ of forests with $n$ edges on $N$ vertices as $N\to\infty$. Here we transliterate in full Britikov's results,\footnote{Britikov's original estimates were phrased in terms of the number $K$ of trees in a forest with $n$ edges on $N$ vertices, which is given by $K=N-n$.} even though we shall essentially use only two.

Let $\nu(N,n)$ be the function defined by
\begin{equation}\label{eq:nu_definition}
\nu(N,n):=N^\frac13\frac{\frac{2n}{N}-1}{\left(1-\frac{n}{N}\right)^\frac23}\quad\textrm{for}\quad n\leqq N-1.
\end{equation}
It is not difficult to show that $\nu$ obeys the bound
\begin{equation}\label{eq:nu_bounds}
\nu\leqq2n-N\leqq N-2.
\end{equation}

Britikov~\cite{britikov:asymtotic} proved that for $N\to\infty$
%\begin{eqnarray}
\begin{equation}\label{eq:Britikov_a}
C(N,n)\approx
{{N}\choose{n}}\left(\frac{N-n}{2}\right)^n \quad\textrm{if}\ n\ \textrm{is bounded}, 
\end{equation}	
\begin{equation}\label{eq:Britikov_b}
C(N,n)\approx\frac{N^{2n}}{2^nn!}\left(1-\frac{2n}{N}\right)^\frac12 \quad\textrm{if}\ n\to\infty\ \textrm{and}\ \nu\to-\infty,
\end{equation}
\begin{eqnarray}\label{eq:Britikov_c}
C(N,n)\approx\frac{N^{N-2}(N-n)^\frac{5}{6}}{2^{N-n-3}(N-n-1)!}&\frac{1}{\sqrt{\pi}}\int_0^\infty\expo{-\frac43t^\frac32}\cos\left(\nu t+\case43 t^\frac32\right)\dd t\nonumber \\
&\textrm{if}\ \nu\ \textrm{is bounded},
\end{eqnarray}
\begin{equation}\label{eq:Britikov_d}
C(N,n)\approx\frac{N^{N-2}}{2^{N-n-1}(N-n-1)!}\left(\frac{2n}{N}-1\right)^{-\frac52} \quad\textrm{if}\ \nu\to\infty.
\end{equation}
It should be noted that by \eqref{eq:nu_bounds} the estimate \eqref{eq:Britikov_d} is only valid for $n\to\infty$. Moreover, by setting $n=N-1$, which corresponds to the case of a forest with a single tree on $N$ vertices, in the limit as $N\to\infty$, \eqref{eq:Britikov_d} delivers Cayley's exact classical result \cite{cayley:theorem},
\begin{equation}\label{eq:Cayley}
C(N,N-1)= N^{N-2}.
\end{equation}

To approximate the sum 
\begin{equation}\label{eq:sum_number_of_states} 
\sum_{n=0}^{N-1}C(N,n)(-1)^n\fA^n
\end{equation}
in \eqref{eq:mean_field_sum_number_of_states} for large values of $N$, we set
\begin{equation}\label{eq:gamma_definition}
n=\floor{\gamma(N-1)},
\end{equation}
where $\floor{\cdot}$ denotes the floor of the enclosed numerical quantity and $\gamma$ is a parameter  in the interval $(0,1]$.\footnote{Similarly, we denote by $\ceil{\cdot}$ the ceiling of a numerical quantity. We recall that, for a real number $a$, $\floor{a}$ is the greatest integer that is smaller than or equal to $a$, whereas $\ceil{a}$ is the smallest integer that is greater than or equal to $a$.} That is,  we cover with straight lines the admissible domain $n\leqq N-1$, where $C(N,n)$ is defined.  Therefore, for given $\gamma$, $\nu$ becomes a function of $N$:
\begin{equation}\label{eq:nu_function_N}
\nu=N^\frac13\frac{2\gamma-1}{(1-\gamma)^\frac23},
\end{equation}
and so only fomulas \eref{eq:Britikov_b}, \eref{eq:Britikov_c}, and \eref{eq:Britikov_d} apply to our case, depending on whether $0<\gamma<\frac12$, $\gamma=\frac12$, or $\frac12<\gamma\leqq1$, respectively.

For $\gamma=\frac12$, \eref{eq:Britikov_c}, which is to be computed for $\nu=0$ and applies only if $\ceil{\frac{N}{2}}=\floor{\frac{N}{2}}$, gives
\begin{equation}\label{eq:C_gamma_1o2}
C\left(N,\frac{N}{2}\right)\approx c_0(\rme N)^\frac{N}{2},
\end{equation}
which was obtained by use of Stirling's approximation in the form
\begin{equation}\label{eq:Stirling_refined}
\left(\frac{N}{2}\right)!\approx\sqrt{2\pi}\left(\frac{N}{2\rme}\right)^\frac{N}{2},
\end{equation}
see \S\,5.11.1 of \cite{NIST:DLMF}. In \eref{eq:C_gamma_1o2}, $c_0$ is a constant that can be evaluated explicitly,
\begin{equation}\label{eq:c_0}
c_0=\int_0^\infty\rme^{-\frac43t^\frac32}\cos\left(\case43t^\frac32\right)\dd t=\frac{1}{16\pi}2^\frac563^\frac16\Gamma\left(\case23\right)\doteq0.512.
\end{equation}
Though we have carefully computed $c_0$, as we shall do below for the similar constants $c_1$ and $c_2$, we are aware that any multiplicative constant scaling $\GNA$ does not affect the free energy per particle $\FNA/N$ in the thermodynamic limit. 

We start by considering $0<\gamma<\frac12$. Making use of \eref{eq:Britikov_b} in \eref{eq:sum_number_of_states}, we  compute 
\begin{equation}\label{eq:sum_number_of_states_A}
\sum_{n=0}^{\ceil{\frac{N}{2}}-1}C(N,n)(-1)^n\fA^n\approx\sum_{n=0}^{\ceil{\frac{N}{2}}-1}\frac{N^{n}}{n!}\left(1-\frac{2n}{N}\right)^\frac12(-x)^n,
\end{equation}
where we have used \eqref{eq:x_definition}.
If we note that for large $N$ the factor $\left(1-\frac{2n}{N}\right)^\frac12$ in \eref{eq:sum_number_of_states_A} varies much more slowly than $\frac{N^{n}}{n!}$, as $n$ increases, we can replace it by its average in the range of variability of $n$. A simple computation shows that
\begin{equation}\label{eq:average_A}
c_1=\ave{\left(1-\frac{2n}{N}\right)^\frac12}=\case13
\end{equation} 
and with the aid of \textsc{maple} \eref{eq:sum_number_of_states_A} can be rewritten as 	
\begin{eqnarray}\label{eq:sum_number_of_states_A_Q}
\sum_{n=0}^{\ceil{\frac{N}{2}}-1}C(N,n)(-1)^n\fA^n&\approx c_1\sum_{n=0}^{\ceil{\frac{N}{2}}-1}\frac{N^{n}}{n!}(-x)^n\nonumber\\ &=Q\left(\ceil{\case{N}{2}},-Nx\right)\expo{-Nx},
\end{eqnarray}	
where $Q$ is the \emph{incomplete} Gamma function \emph{ratio} defined as (\cite{NIST:DLMF}, \S\,8.2.4)
\begin{equation}\label{eq:Q_definition}
Q(a,z):=\frac{\Gamma(a,z)}{\Gamma(a)}.
\end{equation}	
In \eref{eq:Q_definition}, $\Gamma(a)$ and $\Gamma(a,z)$ denote the \emph{incomplete} and \emph{complete} Gamma functions, respectively (\cite{NIST:DLMF}, \S\S\, 8.2.1 and 8.2.2).

We are interested in the behaviour of the sum in \eref{eq:sum_number_of_states_A_Q} for large $N$ and fixed $x$. As shown in \S\,8.11 of \cite{NIST:DLMF}, a number of classic asymptotic formul\ae\ are know for $Q(a,z)$, but in them, invariably, only one argument diverges while the other is kept fixed. Temme~\cite{temme:asymtotic} proved instead  a theorem that describes the asymptotic behaviour of $Q(a,z)$ when both arguments diverge, while their ratio is kept fixed, which is precisely what is needed in \eref{eq:sum_number_of_states_A_Q}. One of Temme's formul\ae\ reads as follows,\footnote{Temme, who had obtained the leading term of \eqref{eq:Temme_formula} in \cite{temme:uniform}, proved in \cite{temme:asymtotic} a recursive formula for the remainder, of which \eqref{eq:Temme_formula} is the first occurrence. Here, we follow, in particular, Sect.~5 of \cite{temme:asymtotic}, where a convention about the choice of the root of \eqref{eq:eta} is given for $\lambda$ ranging in the  complex plane.}
\begin{equation}\label{eq:Temme_formula}
Q(a,z)=\frac12\erfc\left(\eta\sqrt{\frac{a}{2}}\right)+\left(\frac{1}{\lambda-1}-\frac{1}{\eta}\right)\frac{\expo{-\frac12a\eta^2}}{\sqrt{2\pi a}}+\Or\left(\frac{1}{a}\right),
\end{equation}
where $\erfc$ denotes the \emph{complementary} error function (\cite{NIST:DLMF}, \S\,7.2.2),
\begin{equation}\label{eq:lambda}
\lambda=\frac{z}{a}
\end{equation}
is kept fixed as $a\to\infty$, and 
\begin{equation}\label{eq:eta}
\eta=\sqrt{2(\lambda-1-\ln\lambda)}.
\end{equation}
In the case that interests us $\lambda=-2x$, and letting $\eta=\alpha+\rmi\beta$ we convert \eqref{eq:eta} into
\begin{equation}\label{eq:alpha_beta_equations}
\cases{\case12(\alpha^2-\beta^2)=-\mu(x),\\
	\alpha\beta=-\pi}
\end{equation} 
where
\begin{equation}\label{eq:mu}
\mu(x)=1+2x+\ln2x.
\end{equation}

There are two distinct representations of the sum in \eref{eq:sum_number_of_states_A_Q} via \eref{eq:Temme_formula}, depending on the sign of $\mu$. There is a single zero of $\mu(x)$ for $x\geqq0$; this is precisely $x_0$, as defined in \eqref{eq:x_0}.
Also, by use of the asymptotic expression 
\begin{equation}
\erfc\,z\approx\frac{1}{\sqrt{\pi}}\frac{1}{z}\expo{-z^2},\quad z\to\infty,\quad |\mathrm{arg}\,z|<\case{3\pi}{4}
\end{equation}
(\cite{NIST:DLMF}, \S\,7.12.1), and of the symmetry property 
\begin{equation}
\erfc(-z)=2-\erfc\,z
\end{equation} 
(\cite{NIST:DLMF}, \S\,7.4.2), we arrive at 
\begin{eqnarray}
&\sum_{n=0}^{\ceil{\frac{N}{2}-1}}C(N,n)(-1)^n\fA^n\approx\label{eq:partial_sum}\\
&c_1\cases{\expo{-Nx}&for $0<x <x_0$,\\
	(-1)^{\ceil{\frac{N}{2}}+1}\frac{1}{2x+1}\frac{1}{\sqrt{\pi N}}\expo{\left(\ceil{\frac{N}{2}}-\floor{\frac{N}{2}}\right)x}(2\rme x)^\frac{N}{2}&for $x>x_0$.}\label{eq:sum_number_of_states_A_final}
\end{eqnarray}

A key result we obtain from \eref{eq:sum_number_of_states_A_final} is that for $x>x_0$ the partial sum in \eqref{eq:partial_sum} alternates in sign as $N$ grows. Since the average number of states $\GNA$ is a positive quantity, this fact is to be interpreted as a signal that the approximation of setting equal to its maximum the contribution to $\GNA$ from the non-tree graphs in \eref{eq:number_of_states_non_trees} breaks down for $x>x_0$. This alone should suffice to confine our attention to the interval $0<x<x_0$. Nevertheless, we continue our analysis for completeness.

Completing the sum in \eqref{eq:sum_number_of_states} is more complicated, but fortunately it can be estimated via Temme's formula \eqref{eq:Temme_formula}. Here we sketch the essential steps of our reduction strategy. First, we obtain from \eqref{eq:Britikov_d} that 
\begin{eqnarray}
&\sum_{n=\floor{\frac{N}{2}}+1}^{N-1}C(N,n)(-1)^n\fA^n\approx\nonumber\\
&\sum_{n=\floor{\frac{N}{2}}+1}^{N-1}\frac{1}{N}\frac{2^{2n}N^{N-n-1}}{2^{N-1}(N-n-1)!}\left(\frac{2n}{N}-1\right)^{-\frac52}(-x)^n.\label{eq:sum_number_of_states_B}
\end{eqnarray}
Second, we replace the slowly varying function $\left(\frac{2n}{N}-1\right)^{-\frac52}$ with its average in the range $\floor{\frac{N}{2}}+1\leqq n\leqq N-1$, which we estimate through the Euler-Maclaurin formula (\cite{NIST:DLMF}, \S\,2.10.1) as\footnote{Again, as for $c_0$ and $c_1$, the constant $c_2$ is estimated here for completeness; its value will be inessential to the following development.}
\begin{equation}\label{eq:average_B}
\ave{\left(\frac{2n}{N}-1\right)^{-\frac52}}\approx c_2 N^\frac32,\quad c_2=\frac{11}{32}\sqrt{{2}}\doteq0.486.
\end{equation}
Inserting \eref{eq:average_B} into \eref{eq:sum_number_of_states_B} and computing that finite sum with \textsc{maple}, after some algebra we arrive at 
\begin{eqnarray}
&\sum_{n=\floor{\frac{N}{2}}+1}^{N-1}C(N,n)(-1)^n\fA^n\approx\nonumber\\
&c_2\frac{N^{N-\frac12}}{2^{\ceil{\frac{N}{2}}-\floor{\frac{N}{2}}-1}}\frac{\left(-\frac{x}{N}\right)^{\floor{\frac{N}{2}}}}{\Gamma\left(\ceil{\frac{N}{2}}-1\right)}U\left(1,\ceil{\case{N}{2}},-\case{N}{4x}\right),\label{eq:sum_number_of_states_B_U}
\end{eqnarray}
where $U$ is Kummer's confluent hypergeometric function.\footnote{We are guilty of ambiguous notation here: we use the same symbol, $U$, for Kummer's function as for the interaction potential because that is its traditional name. However, the two functions are unlikely to be confused.} Exploiting relationships between $U$ and Whittaker's confluent hypergeometric function (\cite{NIST:DLMF}, \S\,13.14.5) and between this latter and the incomplete Gamma function (\cite{NIST:DLMF}, \S\,13.18.5), one can show that 
\begin{equation}\label{eq:U_and_Gamma}
U\left(1,\ceil{\case{N}{2}},-\case{N}{4x}\right)=\frac{\expo{-\frac{N}{4x}}}{\ceil{\frac{N}{2}}-1}\Gamma\left(\ceil{\case{N}{2}}-1,-\case{N}{4x}\right),
\end{equation}
which combined with \eref{eq:sum_number_of_states_B_U} finally yields
\begin{eqnarray}\label{eq:sum_number_of_states_B_Q}
&\sum_{n=\floor{\frac{N}{2}}+1}^{N-1}C(N,n)(-1)^n\fA^n\approx\nonumber\\
&c_2\sqrt{N}(-2x)^{N-1}\expo{-\frac{N}{4x}}Q\left(\ceil{\case{N}{2}}-1,-\case{N}{4x}\right).
\end{eqnarray}
Reduced to this form, the above sum can be evaluated asymptotically for large $N$ by use of \eqref{eq:Temme_formula}, which here leads us to  
\begin{eqnarray}
%\fl
&
\sum_{n=\floor{\frac{N}{2}}+1}^{N-1}C(N,n)(-1)^n\fA^n\approx\nonumber\\
&
c_2\cases{
	(-1)^{\floor{\frac{N}{2}}+1}\frac{1}{\sqrt{\pi}}\frac{2x}{\rme(2x+1)}\expo{\frac{1}{4x}\left(\ceil{\frac{N}{2}}-\floor{\frac{N}{2}}-2\right)}(2\rme x)^\frac{N}{2}&for $0<x<\frac{1}{4x_0}$,\\
	(-1)^{N+1}\frac{\sqrt{N}}{2x}\left(2x\expo{-\frac{1}{4x}}\right)^N
	&for $x>\frac{1}{4x_0}$,
}\label{eq:sum_number_of_states_B_final}
\end{eqnarray}
where $x_0$ is the same as in \eqref{eq:x_0}.
Both forms of \eref{eq:sum_number_of_states_B_final} are oscillating functions. 

To estimate $\GNA$, since $x_0<\frac{1}{4x_0}$,  we need to put together \eqref{eq:C_gamma_1o2}, \eqref{eq:sum_number_of_states_A_final}, and \eqref{eq:sum_number_of_states_B_final} in the interval $0<x<x_0$, where they  may constitute a positive function. Since by \eqref{eq:x_definition} and \eqref{eq:C_gamma_1o2} the contribution given to the total number of states in \eqref{eq:sum_number_of_states} by the term with $n=\frac{N}{2}$ (for $N$ even) can be written, to leading order, as $(-2\rme x)^\frac{N}{2}$, we arrive at the following estimate
\begin{equation}\label{eq:sum_number_of_states_final}
\GNA\approx V^N\meas^N\expo{-Nx}\left[1+A(x,N)\expo{\frac{N}{2}\mu(x)}\right],\quad 0<x<x_0,
\end{equation}
where an inessential multiplicative constant has been dropped, $A(x,N)$ is a bounded function which can easily be read off from the first line of \eqref{eq:sum_number_of_states_B_final}, and $\mu$ is defined as in \eqref{eq:mu}. Since $\mu(x)<0$ for $0<x<x_0$, it follows from \eqref{eq:sum_number_of_states_final} that for large $N$
\begin{equation}\label{eq:sum_number_of_states_final_estimate}
\GNA\approx V^N\meas^N\expo{-Nx},
\end{equation}
provided that $0<x<x_0$. This is precisely formula \eqref{eq:mean_field_number_of_states} of Sect.~\ref{sec:mean_field}. Our attempts to simplify this lengthy proof of \eqref{eq:sum_number_of_states_final_estimate} have so far failed.

\section*{References}
%\bibliographystyle{iopart-num}
%\bibliography{Missing.bib}
\providecommand{\newblock}{}

\end{document}